\documentclass[11pt,a4paper,english,superscriptaddress,aps]{revtex4}
\usepackage{graphicx}
\makeatletter
\usepackage{babel}
\usepackage[latin1]{inputenc}
\usepackage{hyperref}
\renewcommand{\a}{\alpha}
\renewcommand{\b}{\beta}

\def\m{\mu}

\def\r{\rho}

\def\t{\tau}

\def\x{\xi}

\def\L{\Lambda}

\def\as{a\!\!\!/}
\def\ps{p\!\!\!/}
\def\ks{k\!\!\!/}
\def\bs{b\!\!\!/}

\def\ds{\partial\!\!\!/}

\def\n{\nu}
\def\m{\mu}
\def\n{\nu}

\newcommand{\be}{\begin{equation}}
\newcommand{\ee}{\end{equation}}
\newcommand{\bea}{\begin{eqnarray}}
\newcommand{\eea}{\end{eqnarray}}

\newcommand{\pa}{\partial}

\begin{document}

\immediate\write16{<WARNING: FEYNMAN macros work only with emTeX-dvivers
                    (dviscr.exe, dvihplj.exe, dvidot.exe, etc.) >}
\newdimen\Lengthunit
\newcount\Nhalfperiods
\Lengthunit = 1.5cm
\Nhalfperiods = 9
\catcode`\*=11
\newdimen\L*   \newdimen\d*   \newdimen\d**
\newdimen\dm*  \newdimen\dd*  \newdimen\dt*
\newdimen\a*   \newdimen\b*   \newdimen\c*
\newdimen\a**  \newdimen\b**
\newdimen\xL*  \newdimen\yL*
\newcount\k*   \newcount\l*   \newcount\m*
\newcount\n*   \newcount\dn*  \newcount\r*
\newcount\N*   \newcount\*one \newcount\*two  \*one=1 \*two=2
\newcount\*ths \*ths=1000
\def\GRAPH(hsize=#1)#2{\hbox to #1\Lengthunit{#2\hss}}
\def\Linewidth#1{\special{em:linewidth #1}}
\Linewidth{.4pt}
\def\sm*{\special{em:moveto}}
\def\sl*{\special{em:lineto}}
\newbox\spm*   \newbox\spl*
\setbox\spm*\hbox{\sm*}
\setbox\spl*\hbox{\sl*}
\def\mov#1(#2,#3)#4{\rlap{\L*=#1\Lengthunit\kern#2\L*\raise#3\L*\hbox{#4}}}
\def\smov#1(#2,#3)#4{\rlap{\L*=#1\Lengthunit
\xL*=\xscale\L*\yL*=\yscale\L*\kern#2\xL*\raise#3\yL*\hbox{#4}}}
\def\mov*(#1,#2)#3{\rlap{\kern#1\raise#2\hbox{#3}}}
\def\lin#1(#2,#3){\rlap{\sm*\mov#1(#2,#3){\sl*}}}
\def\arr*(#1,#2,#3){\mov*(#1\dd*,#1\dt*){%
\sm*\mov*(#2\dd*,#2\dt*){\mov*(#3\dt*,-#3\dd*){\sl*}}%
\sm*\mov*(#2\dd*,#2\dt*){\mov*(-#3\dt*,#3\dd*){\sl*}}}}
\def\arrow#1(#2,#3){\rlap{\lin#1(#2,#3)\mov#1(#2,#3){%
\d**=-.012\Lengthunit\dd*=#2\d**\dt*=#3\d**%
\arr*(1,10,4)\arr*(3,8,4)\arr*(4.8,4.2,3)}}}
\def\arrlin#1(#2,#3){\rlap{\L*=#1\Lengthunit\L*=.5\L*%
\lin#1(#2,#3)\mov*(#2\L*,#3\L*){\arrow.1(#2,#3)}}}
\def\dasharrow#1(#2,#3){\rlap{%
{\Lengthunit=0.9\Lengthunit\dashlin#1(#2,#3)\mov#1(#2,#3){\sm*}}%
\mov#1(#2,#3){\sl*\d**=-.012\Lengthunit\dd*=#2\d**\dt*=#3\d**%
\arr*(1,10,4)\arr*(3,8,4)\arr*(4.8,4.2,3)}}}
\def\clap#1{\hbox to 0pt{\hss #1\hss}}
\def\ind(#1,#2)#3{\rlap{%
\d*=.1\Lengthunit\kern#1\d*\raise#2\d*\hbox{\lower2pt\clap{$#3$}}}}
\def\sh*(#1,#2)#3{\rlap{%
\dm*=\the\n*\d**\xL*=\xscale\dm*\yL*=\yscale\dm*
\kern#1\xL*\raise#2\yL*\hbox{#3}}}
\def\calcnum*#1(#2,#3){\a*=1000sp\b*=1000sp\a*=#2\a*\b*=#3\b*%
\ifdim\a*<0pt\a*-\a*\fi\ifdim\b*<0pt\b*-\b*\fi%
\ifdim\a*>\b*\c*=.96\a*\advance\c*.4\b*%
\else\c*=.96\b*\advance\c*.4\a*\fi%
\k*\a*\multiply\k*\k*\l*\b*\multiply\l*\l*%
\m*\k*\advance\m*\l*\n*\c*\r*\n*\multiply\n*\n*%
\dn*\m*\advance\dn*-\n*\divide\dn*2\divide\dn*\r*%
\advance\r*\dn*%
\c*=\the\Nhalfperiods5sp\c*=#1\c*\ifdim\c*<0pt\c*-\c*\fi%
\multiply\c*\r*\N*\c*\divide\N*10000}
\def\dashlin#1(#2,#3){\rlap{\calcnum*#1(#2,#3)%
\d**=#1\Lengthunit\ifdim\d**<0pt\d**-\d**\fi%
\divide\N*2\multiply\N*2\advance\N*1%
\divide\d**\N*\sm*\n*\*one\sh*(#2,#3){\sl*}%
\loop\advance\n*\*one\sh*(#2,#3){\sm*}\advance\n*\*one\sh*(#2,#3){\sl*}%
\ifnum\n*<\N*\repeat}}
\def\dashdotlin#1(#2,#3){\rlap{\calcnum*#1(#2,#3)%
\d**=#1\Lengthunit\ifdim\d**<0pt\d**-\d**\fi%
\divide\N*2\multiply\N*2\advance\N*1\multiply\N*2%
\divide\d**\N*\sm*\n*\*two\sh*(#2,#3){\sl*}\loop%
\advance\n*\*one\sh*(#2,#3){\kern-1.48pt\lower.5pt\hbox{\rm.}}%
\advance\n*\*one\sh*(#2,#3){\sm*}%
\advance\n*\*two\sh*(#2,#3){\sl*}\ifnum\n*<\N*\repeat}}
\def\shl*(#1,#2)#3{\kern#1#3\lower#2#3\hbox{\unhcopy\spl*}}
\def\trianglin#1(#2,#3){\rlap{\toks0={#2}\toks1={#3}\calcnum*#1(#2,#3)%
\dd*=.57\Lengthunit\dd*=#1\dd*\divide\dd*\N*%
\d**=#1\Lengthunit\ifdim\d**<0pt\d**-\d**\fi%
\multiply\N*2\divide\d**\N*\advance\N*-1\sm*\n*\*one\loop%
\shl**{\dd*}\dd*-\dd*\advance\n*2%
\ifnum\n*<\N*\repeat\n*\N*\advance\n*1\shl**{0pt}}}
\def\wavelin#1(#2,#3){\rlap{\toks0={#2}\toks1={#3}\calcnum*#1(#2,#3)%
\dd*=.23\Lengthunit\dd*=#1\dd*\divide\dd*\N*%
\d**=#1\Lengthunit\ifdim\d**<0pt\d**-\d**\fi%
\multiply\N*4\divide\d**\N*\sm*\n*\*one\loop%
\shl**{\dd*}\dt*=1.3\dd*\advance\n*1%
\shl**{\dt*}\advance\n*\*one%
\shl**{\dd*}\advance\n*\*two%
\dd*-\dd*\ifnum\n*<\N*\repeat\n*\N*\shl**{0pt}}}
\def\w*lin(#1,#2){\rlap{\toks0={#1}\toks1={#2}\d**=\Lengthunit\dd*=-.12\d**%
\N*8\divide\d**\N*\sm*\n*\*one\loop%
\shl**{\dd*}\dt*=1.3\dd*\advance\n*\*one%
\shl**{\dt*}\advance\n*\*one%
\shl**{\dd*}\advance\n*\*one%
\shl**{0pt}\dd*-\dd*\advance\n*1\ifnum\n*<\N*\repeat}}
\def\l*arc(#1,#2)[#3][#4]{\rlap{\toks0={#1}\toks1={#2}\d**=\Lengthunit%
\dd*=#3.037\d**\dd*=#4\dd*\dt*=#3.049\d**\dt*=#4\dt*\ifdim\d**>16mm%
\d**=.25\d**\n*\*one\shl**{-\dd*}\n*\*two\shl**{-\dt*}\n*3\relax%
\shl**{-\dd*}\n*4\relax\shl**{0pt}\else\ifdim\d**>5mm%
\d**=.5\d**\n*\*one\shl**{-\dt*}\n*\*two\shl**{0pt}%
\else\n*\*one\shl**{0pt}\fi\fi}}
\def\d*arc(#1,#2)[#3][#4]{\rlap{\toks0={#1}\toks1={#2}\d**=\Lengthunit%
\dd*=#3.037\d**\dd*=#4\dd*\d**=.25\d**\sm*\n*\*one\shl**{-\dd*}%
\n*3\relax\sh*(#1,#2){\xL*=\xscale\dd*\yL*=\yscale\dd*
\kern#2\xL*\lower#1\yL*\hbox{\sm*}}%
\n*4\relax\shl**{0pt}}}
\def\arc#1[#2][#3]{\rlap{\Lengthunit=#1\Lengthunit%
\sm*\l*arc(#2.1914,#3.0381)[#2][#3]%
\smov(#2.1914,#3.0381){\l*arc(#2.1622,#3.1084)[#2][#3]}%
\smov(#2.3536,#3.1465){\l*arc(#2.1084,#3.1622)[#2][#3]}%
\smov(#2.4619,#3.3086){\l*arc(#2.0381,#3.1914)[#2][#3]}}}
\def\dasharc#1[#2][#3]{\rlap{\Lengthunit=#1\Lengthunit%
\d*arc(#2.1914,#3.0381)[#2][#3]%
\smov(#2.1914,#3.0381){\d*arc(#2.1622,#3.1084)[#2][#3]}%
\smov(#2.3536,#3.1465){\d*arc(#2.1084,#3.1622)[#2][#3]}%
\smov(#2.4619,#3.3086){\d*arc(#2.0381,#3.1914)[#2][#3]}}}
\def\wavearc#1[#2][#3]{\rlap{\Lengthunit=#1\Lengthunit%
\w*lin(#2.1914,#3.0381)%
\smov(#2.1914,#3.0381){\w*lin(#2.1622,#3.1084)}%
\smov(#2.3536,#3.1465){\w*lin(#2.1084,#3.1622)}%
\smov(#2.4619,#3.3086){\w*lin(#2.0381,#3.1914)}}}
\def\shl**#1{\c*=\the\n*\d**\d*=#1%
\a*=\the\toks0\c*\b*=\the\toks1\d*\advance\a*-\b*%
\b*=\the\toks1\c*\d*=\the\toks0\d*\advance\b*\d*%
\a*=\xscale\a*\b*=\yscale\b*%
\raise\b*\rlap{\kern\a*\unhcopy\spl*}}
\def\wlin*#1(#2,#3)[#4]{\rlap{\toks0={#2}\toks1={#3}%
\c*=#1\l*\c*\c*=.01\Lengthunit\m*\c*\divide\l*\m*%
\c*=\the\Nhalfperiods5sp\multiply\c*\l*\N*\c*\divide\N*\*ths%
\divide\N*2\multiply\N*2\advance\N*1%
\dd*=.002\Lengthunit\dd*=#4\dd*\multiply\dd*\l*\divide\dd*\N*%
\d**=#1\multiply\N*4\divide\d**\N*\sm*\n*\*one\loop%
\shl**{\dd*}\dt*=1.3\dd*\advance\n*\*one%
\shl**{\dt*}\advance\n*\*one%
\shl**{\dd*}\advance\n*\*two%
\dd*-\dd*\ifnum\n*<\N*\repeat\n*\N*\shl**{0pt}}}
\def\wavebox#1{\setbox0\hbox{#1}%
\a*=\wd0\advance\a*14pt\b*=\ht0\advance\b*\dp0\advance\b*14pt%
\hbox{\kern9pt%
\mov*(0pt,\ht0){\mov*(-7pt,7pt){\wlin*\a*(1,0)[+]\wlin*\b*(0,-1)[-]}}%
\mov*(\wd0,-\dp0){\mov*(7pt,-7pt){\wlin*\a*(-1,0)[+]\wlin*\b*(0,1)[-]}}%
\box0\kern9pt}}
\def\rectangle#1(#2,#3){%
\lin#1(#2,0)\lin#1(0,#3)\mov#1(0,#3){\lin#1(#2,0)}\mov#1(#2,0){\lin#1(0,#3)}}
\def\dashrectangle#1(#2,#3){\dashlin#1(#2,0)\dashlin#1(0,#3)%
\mov#1(0,#3){\dashlin#1(#2,0)}\mov#1(#2,0){\dashlin#1(0,#3)}}
\def\waverectangle#1(#2,#3){\L*=#1\Lengthunit\a*=#2\L*\b*=#3\L*%
\ifdim\a*<0pt\a*-\a*\def\x*{-1}\else\def\x*{1}\fi%
\ifdim\b*<0pt\b*-\b*\def\y*{-1}\else\def\y*{1}\fi%
\wlin*\a*(\x*,0)[-]\wlin*\b*(0,\y*)[+]%
\mov#1(0,#3){\wlin*\a*(\x*,0)[+]}\mov#1(#2,0){\wlin*\b*(0,\y*)[-]}}
\def\calcparab*{%
\ifnum\n*>\m*\k*\N*\advance\k*-\n*\else\k*\n*\fi%
\a*=\the\k* sp\a*=10\a*\b*\dm*\advance\b*-\a*\k*\b*%
\a*=\the\*ths\b*\divide\a*\l*\multiply\a*\k*%
\divide\a*\l*\k*\*ths\r*\a*\advance\k*-\r*%
\dt*=\the\k*\L*}
\def\arcto#1(#2,#3)[#4]{\rlap{\toks0={#2}\toks1={#3}\calcnum*#1(#2,#3)%
\dm*=135sp\dm*=#1\dm*\d**=#1\Lengthunit\ifdim\dm*<0pt\dm*-\dm*\fi%
\multiply\dm*\r*\a*=.3\dm*\a*=#4\a*\ifdim\a*<0pt\a*-\a*\fi%
\advance\dm*\a*\N*\dm*\divide\N*10000%
\divide\N*2\multiply\N*2\advance\N*1%
\L*=-.25\d**\L*=#4\L*\divide\d**\N*\divide\L*\*ths%
\m*\N*\divide\m*2\dm*=\the\m*5sp\l*\dm*%
\sm*\n*\*one\loop\calcparab*\shl**{-\dt*}%
\advance\n*1\ifnum\n*<\N*\repeat}}
\def\arrarcto#1(#2,#3)[#4]{\L*=#1\Lengthunit\L*=.54\L*%
\arcto#1(#2,#3)[#4]\mov*(#2\L*,#3\L*){\d*=.457\L*\d*=#4\d*\d**-\d*%
\mov*(#3\d**,#2\d*){\arrow.02(#2,#3)}}}
\def\dasharcto#1(#2,#3)[#4]{\rlap{\toks0={#2}\toks1={#3}\calcnum*#1(#2,#3)%
\dm*=\the\N*5sp\a*=.3\dm*\a*=#4\a*\ifdim\a*<0pt\a*-\a*\fi%
\advance\dm*\a*\N*\dm*%
\divide\N*20\multiply\N*2\advance\N*1\d**=#1\Lengthunit%
\L*=-.25\d**\L*=#4\L*\divide\d**\N*\divide\L*\*ths%
\m*\N*\divide\m*2\dm*=\the\m*5sp\l*\dm*%
\sm*\n*\*one\loop%
\calcparab*\shl**{-\dt*}\advance\n*1%
\ifnum\n*>\N*\else\calcparab*%
\sh*(#2,#3){\kern#3\dt*\lower#2\dt*\hbox{\sm*}}\fi%
\advance\n*1\ifnum\n*<\N*\repeat}}
\def\*shl*#1{%
\c*=\the\n*\d**\advance\c*#1\a**\d*\dt*\advance\d*#1\b**%
\a*=\the\toks0\c*\b*=\the\toks1\d*\advance\a*-\b*%
\b*=\the\toks1\c*\d*=\the\toks0\d*\advance\b*\d*%
\raise\b*\rlap{\kern\a*\unhcopy\spl*}}
\def\calcnormal*#1{%
\b**=10000sp\a**\b**\k*\n*\advance\k*-\m*%
\multiply\a**\k*\divide\a**\m*\a**=#1\a**\ifdim\a**<0pt\a**-\a**\fi%
\ifdim\a**>\b**\d*=.96\a**\advance\d*.4\b**%
\else\d*=.96\b**\advance\d*.4\a**\fi%
\d*=.01\d*\r*\d*\divide\a**\r*\divide\b**\r*%
\ifnum\k*<0\a**-\a**\fi\d*=#1\d*\ifdim\d*<0pt\b**-\b**\fi%
\k*\a**\a**=\the\k*\dd*\k*\b**\b**=\the\k*\dd*}
\def\wavearcto#1(#2,#3)[#4]{\rlap{\toks0={#2}\toks1={#3}\calcnum*#1(#2,#3)%
\c*=\the\N*5sp\a*=.4\c*\a*=#4\a*\ifdim\a*<0pt\a*-\a*\fi%
\advance\c*\a*\N*\c*\divide\N*20\multiply\N*2\advance\N*-1\multiply\N*4%
\d**=#1\Lengthunit\dd*=.012\d**\ifdim\d**<0pt\d**-\d**\fi\L*=.25\d**%
\divide\d**\N*\divide\dd*\N*\L*=#4\L*\divide\L*\*ths%
\m*\N*\divide\m*2\dm*=\the\m*0sp\l*\dm*%
\sm*\n*\*one\loop\calcnormal*{#4}\calcparab*%
\*shl*{1}\advance\n*\*one\calcparab*%
\*shl*{1.3}\advance\n*\*one\calcparab*%
\*shl*{1}\advance\n*2%
\dd*-\dd*\ifnum\n*<\N*\repeat\n*\N*\shl**{0pt}}}
\def\triangarcto#1(#2,#3)[#4]{\rlap{\toks0={#2}\toks1={#3}\calcnum*#1(#2,#3)%
\c*=\the\N*5sp\a*=.4\c*\a*=#4\a*\ifdim\a*<0pt\a*-\a*\fi%
\advance\c*\a*\N*\c*\divide\N*20\multiply\N*2\advance\N*-1\multiply\N*2%
\d**=#1\Lengthunit\dd*=.012\d**\ifdim\d**<0pt\d**-\d**\fi\L*=.25\d**%
\divide\d**\N*\divide\dd*\N*\L*=#4\L*\divide\L*\*ths%
\m*\N*\divide\m*2\dm*=\the\m*0sp\l*\dm*%
\sm*\n*\*one\loop\calcnormal*{#4}\calcparab*%
\*shl*{1}\advance\n*2%
\dd*-\dd*\ifnum\n*<\N*\repeat\n*\N*\shl**{0pt}}}
\def\hr*#1{\clap{\xL*=\xscale\Lengthunit\vrule width#1\xL* height.1pt}}
\def\shade#1[#2]{\rlap{\Lengthunit=#1\Lengthunit%
\smov(0,#2.05){\hr*{.994}}\smov(0,#2.1){\hr*{.980}}%
\smov(0,#2.15){\hr*{.953}}\smov(0,#2.2){\hr*{.916}}%
\smov(0,#2.25){\hr*{.867}}\smov(0,#2.3){\hr*{.798}}%
\smov(0,#2.35){\hr*{.715}}\smov(0,#2.4){\hr*{.603}}%
\smov(0,#2.45){\hr*{.435}}}}
\def\dshade#1[#2]{\rlap{%
\Lengthunit=#1\Lengthunit\if#2-\def\t*{+}\else\def\t*{-}\fi%
\smov(0,\t*.025){%
\smov(0,#2.05){\hr*{.995}}\smov(0,#2.1){\hr*{.988}}%
\smov(0,#2.15){\hr*{.969}}\smov(0,#2.2){\hr*{.937}}%
\smov(0,#2.25){\hr*{.893}}\smov(0,#2.3){\hr*{.836}}%
\smov(0,#2.35){\hr*{.760}}\smov(0,#2.4){\hr*{.662}}%
\smov(0,#2.45){\hr*{.531}}\smov(0,#2.5){\hr*{.320}}}}}
\def\vdot{\rlap{\kern-1.9pt\lower1.8pt\hbox{$\scriptstyle\bullet$}}}
\def\vtimes{\rlap{\kern-3pt\lower1.8pt\hbox{$\scriptstyle\times$}}}
\def\vDot{\rlap{\kern-2.3pt\lower2.7pt\hbox{$\bullet$}}}
\def\vTimes{\rlap{\kern-3.6pt\lower2.4pt\hbox{$\times$}}}
\catcode`\*=12
\newcount\CatcodeOfAtSign
\CatcodeOfAtSign=\the\catcode`\@
\catcode`\@=11
\newcount\n@ast
\def\n@ast@#1{\n@ast0\relax\get@ast@#1\end}
\def\get@ast@#1{\ifx#1\end\let\next\relax\else%
\ifx#1*\advance\n@ast1\fi\let\next\get@ast@\fi\next}
\newif\if@up \newif\if@dwn
\def\up@down@#1{\@upfalse\@dwnfalse%
\if#1u\@uptrue\fi\if#1U\@uptrue\fi\if#1+\@uptrue\fi%
\if#1d\@dwntrue\fi\if#1D\@dwntrue\fi\if#1-\@dwntrue\fi}
\def\halfcirc#1(#2)[#3]{{\Lengthunit=#2\Lengthunit\up@down@{#3}%
\if@up\smov(0,.5){\arc[-][-]\arc[+][-]}\fi%
\if@dwn\smov(0,-.5){\arc[-][+]\arc[+][+]}\fi%
\def\lft{\smov(0,.5){\arc[-][-]}\smov(0,-.5){\arc[-][+]}}%
\def\rght{\smov(0,.5){\arc[+][-]}\smov(0,-.5){\arc[+][+]}}%
\if#3l\lft\fi\if#3L\lft\fi\if#3r\rght\fi\if#3R\rght\fi%
\n@ast@{#1}%
\ifnum\n@ast>0\if@up\shade[+]\fi\if@dwn\shade[-]\fi\fi%
\ifnum\n@ast>1\if@up\dshade[+]\fi\if@dwn\dshade[-]\fi\fi}}
\def\halfdashcirc(#1)[#2]{{\Lengthunit=#1\Lengthunit\up@down@{#2}%
\if@up\smov(0,.5){\dasharc[-][-]\dasharc[+][-]}\fi%
\if@dwn\smov(0,-.5){\dasharc[-][+]\dasharc[+][+]}\fi%
\def\lft{\smov(0,.5){\dasharc[-][-]}\smov(0,-.5){\dasharc[-][+]}}%
\def\rght{\smov(0,.5){\dasharc[+][-]}\smov(0,-.5){\dasharc[+][+]}}%
\if#2l\lft\fi\if#2L\lft\fi\if#2r\rght\fi\if#2R\rght\fi}}
\def\halfwavecirc(#1)[#2]{{\Lengthunit=#1\Lengthunit\up@down@{#2}%
\if@up\smov(0,.5){\wavearc[-][-]\wavearc[+][-]}\fi%
\if@dwn\smov(0,-.5){\wavearc[-][+]\wavearc[+][+]}\fi%
\def\lft{\smov(0,.5){\wavearc[-][-]}\smov(0,-.5){\wavearc[-][+]}}%
\def\rght{\smov(0,.5){\wavearc[+][-]}\smov(0,-.5){\wavearc[+][+]}}%
\if#2l\lft\fi\if#2L\lft\fi\if#2r\rght\fi\if#2R\rght\fi}}
\def\Circle#1(#2){\halfcirc#1(#2)[u]\halfcirc#1(#2)[d]\n@ast@{#1}%
\ifnum\n@ast>0\clap{%
\dimen0=\xscale\Lengthunit\vrule width#2\dimen0 height.1pt}\fi}
\def\wavecirc(#1){\halfwavecirc(#1)[u]\halfwavecirc(#1)[d]}
\def\dashcirc(#1){\halfdashcirc(#1)[u]\halfdashcirc(#1)[d]}
%
\def\xscale{1}
\def\yscale{1}
\def\Ellipse#1(#2)[#3,#4]{\def\xscale{#3}\def\yscale{#4}%
\Circle#1(#2)\def\xscale{1}\def\yscale{1}}
\def\dashEllipse(#1)[#2,#3]{\def\xscale{#2}\def\yscale{#3}%
\dashcirc(#1)\def\xscale{1}\def\yscale{1}}
\def\waveEllipse(#1)[#2,#3]{\def\xscale{#2}\def\yscale{#3}%
\wavecirc(#1)\def\xscale{1}\def\yscale{1}}
\def\halfEllipse#1(#2)[#3][#4,#5]{\def\xscale{#4}\def\yscale{#5}%
\halfcirc#1(#2)[#3]\def\xscale{1}\def\yscale{1}}
\def\halfdashEllipse(#1)[#2][#3,#4]{\def\xscale{#3}\def\yscale{#4}%
\halfdashcirc(#1)[#2]\def\xscale{1}\def\yscale{1}}
\def\halfwaveEllipse(#1)[#2][#3,#4]{\def\xscale{#3}\def\yscale{#4}%
\halfwavecirc(#1)[#2]\def\xscale{1}\def\yscale{1}}
\catcode`\@=\the\CatcodeOfAtSign

\title{On the aether-like Lorentz-breaking actions}

\author{M. Gomes}
\affiliation{Instituto de F\'\i sica, Universidade de S\~ao Paulo\\
Caixa Postal 66318, 05315-970, S\~ao Paulo, SP, Brazil}
\email{mgomes,ajsilva@fma.if.usp.br}

\author{J. R. Nascimento}

\affiliation{Departamento de F\'{\i}sica, Universidade Federal da 
Para\'{\i}ba\\
 Caixa Postal 5008, 58051-970, Jo\~ao Pessoa, Para\'{\i}ba, Brazil}
\email{jroberto,petrov@fisica.ufpb.br}

\author{A. Yu. Petrov}

\affiliation{Departamento de F\'{\i}sica, Universidade Federal da 
Para\'{\i}ba\\
 Caixa Postal 5008, 58051-970, Jo\~ao Pessoa, Para\'{\i}ba, Brazil}
\email{jroberto,petrov@fisica.ufpb.br}

\author{A. J. da Silva}
\affiliation{Instituto de F\'\i sica, Universidade de S\~ao Paulo\\
Caixa Postal 66318, 05315-970, S\~ao Paulo, SP, Brazil}
\email{mgomes,ajsilva@fma.if.usp.br}

\begin{abstract}
We show that CPT-even aether-like Lorentz-breaking actions,
for the scalar and electromagnetic fields, are generated 
via their appropriate Lorentz-breaking coupling to spinor fields, in
three, four and five
space-time dimensions. Besides, we also show that aether-like terms for
the spinor field can be generated as a consequence of the same
couplings. We discuss the dispersion relations in the theories
with  aether-like Lorentz-breaking terms and find the tree-level
effective  (Breit) potential for fermion
scattering and the one-loop effective potential corresponding to the
action  of the scalar field.
\end{abstract}

\maketitle
\newpage
\section{Introduction}
The possibility of Lorentz symmetry breaking \cite{JK} has attracted a
great  deal of
 attention during the recent years. One of the most important
directions in this study is the investigation of  possible
Lorentz-violating extensions of  field theory models. A general
review of acceptable forms of such extensions was presented in
\cite{Kostel}. The most known examples of  Lorentz-breaking
modifications of  field theories are CPT-odd since the Lorentz
symmetry breaking in these models is implemented through a constant
vector field which is known to break the CPT symmetry \cite{KostColl}. The
well known examples of  Lorentz  and CPT breaking terms are the
Carroll-Field-Jackiw term in electrodynamics \cite{list},  its non-Abelian
generalization \cite{YM} and the gravitational Chern-Simons term
\cite{gra}. Also, the Lorentz-CPT breaking terms which deserve to be
mentioned are the Chern-Simons like term for the scalar fields (in two 
dimensions)
\cite{passos} and the one-derivative Chern-Simons like term in the
linearized  gravity \cite{ourgra}.

However, the CPT-odd terms do not exhaust the possibilities for 
Lorentz-breaking terms. Alternative types of Lorentz-breaking
terms are CPT-even ones. Some of their forms were originally
introduced in \cite{KostColl2} within the context of the
Lorentz-breaking extension of the standard model. Recently, the
interest in such terms was increased due to the development of the aether
concept \cite{aether} and to the study of extra dimensions
\cite{Cleaver}.  Therefore, a
natural problem is the study of the generation and physical impacts of the
aether-like terms, which can be  treated as the simplest
examples of  CPT-even Lorentz-breaking terms \cite{Carroll}. Some
aspects related to such terms, including the study of the dispersion
relations in theories involving CPT-even Lorentz-breaking terms, the  
generation of such
terms via  the gauge embedding method, impacts of these terms at
finite temperature and some experimental estimations for such terms,
were considered  in \cite{even}. Also, in the papers \cite{Carroll},
the modifications of dispersion relations of the scalar, spinor and
electromagnetic fields due to CPT-even couplings of these fields with
the aether field were studied. 
 
In this paper, we show that the CPT-even couplings of the scalar,
spinor and electromagnetic fields with the aether field proposed in
\cite{Carroll} can arise in three, four and  five-dimensional
space-times  as radiative corrections generated by appropriate CPT-odd
Lorentz-breaking interactions of these fields between themselves and
with a constant aether field. We also discuss their physical impacts,
coming from the analysis of  dispersion relations and the effective  potential.

\section{Aether terms in the scalar field theory}
We start with the model of a spinor field coupled to a scalar
matter in a  Lorentz-breaking manner:
\bea
\label{ac1}
S=\int d^Dx\Big[\bar{\psi}(i\ds-m)\psi+\frac{1}{2}(\pa^b\phi\pa_b\phi+
m^2\phi^2)-g\bar{\psi}\as\psi\phi\Big].
\eea
Here $a^b$ is a constant vector implementing the Lorentz symmetry
breaking. The coupling introduced in this way is a natural
Lorentz-breaking generalization of the Yukawa coupling. We consider
this model in three, four and five dimensional space-times. 
The Lorentz-breaking coupling in (\ref{ac1}) is similar to the
general structure of Eq. (6) of Ref. \cite{KostColl2}, with the 
identification $g\bar{\psi}\as\psi\phi=-[(G_L)_{AB}\bar{L}_A\phi
R_B+(G_U)_{AB}\bar{Q}_A\phi^c U_B+(G_D)_{AB}\bar{Q}_A\phi D_B]+h.c.$;
notice however that in our case there is only
one spinor (not chiral) field and no gauge interaction. 
As $\phi$ is real, we have $\phi^c=\phi$.

In this theory, the following Feynman rules take place:

\vspace*{2mm}

\hspace{1cm}
\Lengthunit=1.2cm
\GRAPH(hsize=3){\Linewidth{.6pt}\dashlin(1,0)\ind(15,0){\;\;\;\;\;\;\;\;\;\;\;
    \;=\frac{i(\ps+m)}{p^{2}-m^{2}}}
}
\hspace{1cm}
\Lengthunit=1.2cm
\GRAPH(hsize=3){\ind(5,0){\bullet}\dashlin(1,0)\mov(.5,0){\lin(0,.7)}
  \ind(18,0){\;\;\;\;\;\;\;\;\;=-g\as,}
}
\hspace{1cm}
\Lengthunit=1.2cm
\GRAPH(hsize=3){\Linewidth{.6pt}\lin(1,0)\ind(15,0){\;\;\;\;\;\;\;\;\;\;\;
    \;=\frac{i}{p^{2}-m^{2}}}
}

\vspace*{2mm}

\noindent Thus,  the lowest order contribution to the two-point vertex 
function of the scalar field is given by the following diagram:

\vspace*{3mm}

\hspace{6.0cm}
\Lengthunit=1.2cm
\GRAPH(hsize=2){\lin(.5,0)\mov(1,0){\dashcirc(1)}\mov(1.5,0){\lin(.5,0)}
 \ind(14,0){\bullet}\ind(4,0){\bullet}
}

\vspace*{3mm}

\noindent Here, the simple line is for the external $\phi$ field, and
the dashed  line for the propagator of the spinor $\psi$ field.
This diagram produces the following contribution to the effective action:
\bea\label{t1}
S_2(p)&=&\frac{g^2}{2}\phi(p)\phi(-p)\int\frac{d^Dk}{(2\pi)^D}{\rm tr}
\bigl[\as\,S(k)\as\,S(k+p)].
=\nonumber\\&=&-\frac{g^2}{2}\phi(p)\phi(-p)\int\frac{d^Dk}{(2\pi)^D}{\rm
  tr}\big[\as(\ks +m)\as(\ks+\ps+m)\big]\frac{1}{[k^2-m^2][(k+p)^2-m^2]}.
\eea
 The key problem now consists in calculating the matrix
traces. First of all,  in five-dimensional space
the gamma matrices are $4\times 4$ with $\gamma^0\ldots\gamma^3$ being
the same as in four dimensions, and the five-dimensional $\gamma^4$
coinciding with the four-dimensional chirality matrix $\gamma_5$
(indeed, $\gamma_5$ anticommutes with each $\gamma^a$, $a=0...3$, and
$(\gamma_5)^2=-{\bf 1}$). The gamma matrices defined in this way
satisfy the definition $\{\gamma^a,\gamma^b\}=2\eta^{ab}$, with
$\eta^{ab}= diag(+-\ldots -)$. In three-dimensional space the gamma
matrices are $(\gamma^0)^{\alpha}_{\phantom a\beta}=\sigma^2$,
$(\gamma^1)^{\alpha}_{\phantom a\beta}= i\sigma^1$, 
$(\gamma^2)^{\alpha}_{\phantom a\beta}=i\sigma^3$. 

 We can verify that the well known four-dimensional relation for
the trace of the product $\gamma^a\gamma^b\gamma^c\gamma^d$, can be generalized
also at least in three and five space-time dimensions as
\bea
\label{trg}
&&{\rm
  tr}(\gamma^a\gamma^b\gamma^c\gamma^d)=d(\eta^{ab}\eta^{cd}-\eta^{ac}
\eta^{bd}+\eta^{ad}\eta^{bc}),\nonumber\\
&&{\rm tr}(\gamma^a\gamma^b)=d\eta^{ab}.
\eea
where $d$ is the dimension of the gamma matrix in the corresponding space-time.
The trace of the product of three gamma matrices cannot produce an
aether-like term giving either zero (in four or five dimensions)  or the
Levi-Civita symbol (in three dimensions), and
$a^ma_m=a^2$ which does not generate a Lorentz-breaking
term. Thus,  omitting the $a^2$ terms, we find the only Lorentz
breaking  contribution from (\ref{t1}):
\bea\label{i2}
S_2(p)\simeq -\frac{d}{2}g^2\phi(p)\phi(-p)[\eta^{ab}\eta^{cd}+
\eta^{ad}
\eta^{bc}]a_aa_c\int\frac{d^Dk}{(2\pi)^D}\frac{k_b(k_d+p_d)}{[k^2-
  m^2][(k+p)^2 -m^2]}.
\eea
Then, one may use the Feynman representation with the parameter
$x$ which yields
\bea\label{i3}
S_2(p)\simeq
-dg^2\phi(p)\phi(-p)a^ba^d\int\frac{d^Dk}{(2\pi)^D}\int_0^1 dx
\frac{k_bk_d- p_bp_dx(1-x)}{[k^2-m^2+p^2x(1-x)]^2}.
\eea
The term proportional to $k_bk_d$ after integration gives $\eta_{bd}$,
thus, the corresponding term in the effective action will be
proportional to $a^2$ which does not break the Lorentz symmetry. The
desired 
Lorentz-breaking correction, after Wick rotation, takes the form
\bea\label{i4}
S_2(p)\simeq -idg^2\phi(p)\phi(-p)(a\cdot p)^
2\int\frac{d^Dk_E}{(2\pi)^D} \int_0^1 dxx(1-x)\frac{1}{[k^2+m^2+p^2x(1-x)]^2}.
\eea
We find that in five dimensions this contribution is finite only within the
framework of the dimensional regularization:
\bea
S_2^{D=5}(p)&=&\frac{ig^2|m|}{24\pi^2}\phi(p)\phi(-p)(a\cdot p)^2.
\eea
Thus, we  conclude that in five dimensions the desired finite
aether-like  term \cite{aether} of the form
\bea
\label{aether}
S^{D=5}_{aether}=\frac{1}{M}\int d^5x\phi(a\cdot \pa)^2\phi
\eea
is generated, with $\frac{1}{M}=\frac{|m|g^2}{24\pi^2}$. The
parameters $a^b$ and $M$ are related with the parameter
$(k_{\phi\phi})^{bc}$ from \cite{KostColl2} as
$\frac{1}{2}(k_{\phi\phi})^{bc}=-\frac{a^ba^c}{M}$.

In three space-time dimensions the aether-like term is explicitly
finite without any regularization being also of the form
(\ref{aether}), but with $\frac{1}{M}=\frac{g^2}{16\pi|m|}$.

Finally, in four space-time dimensions we find
\bea
\label{aether4}
S^{D=4}_{aether}=\frac{g^2}{12\pi^2\epsilon}(1-\frac{\epsilon}{2}\ln
\frac{m^2}{\mu^2})\int d^4x\phi(a\cdot \pa)^2\phi,
\eea
where the  divergence  can only be removed  by adding a
counterterm of the form
$S_{aether}^{ct}=-\frac{g^2}{12\pi^2\epsilon}\int d^4x \phi(a\cdot
\pa)^2\phi$.  Thus, first, we find that the aether term should be present
in the theory from the very beginning to provide a consistent one-loop
renormalization and  second, the renormalized one-loop aether contribution
looks like
\bea
\label{aether4a}
S^{D=4}_{aether}=-\frac{g^2}{24\pi^2}\ln\frac{m^2}{\mu^2}
\int d^4x\phi(a\cdot \pa)^2\phi,
\eea
which reproduces the desired aether-like structure, 
with $\frac{1}{M}=-\frac{g^2}{24\pi^2}\ln\frac{m^2}{\mu^2}$. We note that in
four dimensions the aether-like contribution is renormalizable by
dimensional reasons.

After including the
additive aether term (\ref{aether}), the action of the scalar field
takes 
the form
\bea
\label{acae}
S=-\frac{1}{2}\int d^Dx\phi[\pa_b\pa^b+m^2+\frac{1}{M}(a\cdot\pa)^2]\phi,
\eea
which implies in the following dispersion relations:

(i) Time-like $a^b$: $E=\pm\sqrt{\frac{\vec{k}^
    2+m^2}{1+\frac{a^2_0}{M}}}$.  In this case there is no
restrictions on the  dynamics.

(ii) Space-like $a^b$: $E=\pm\sqrt{\vec{k}^2+m^2-
\frac{(\vec{a}\cdot\vec{k})^2}{M}}$. In this case the dynamics is well
defined only at small $|\vec{a}|$.

The propagator of this model has the form
\bea
G(k)=-\frac{1}{k^2-m^2+\frac{1}{M}(a\cdot k)^2}.
\eea
Following \cite{Alts}, we can find the tree-level Breit potential corresponding to fermion-fermion scattering in the
theory where fermions are coupled via Yukawa interaction to a scalar
fields with this propagator. Indeed, in the non-relativistic limit the
incoming (outcoming) fermions with mass $M$ have initial (final)
momenta $p$ and $k$ (respectively $p'$ and $k'$) whose explicit form
is $p=(M,\vec{p})$, $k=(M,\vec{k})$ etc. The momentum exchanged is  
$q=p'-p=(0,\vec{p}'-\vec{p})$.

Thus, the matrix element for the scattering process is (cf. \cite{Alts})
\bea
i{\cal M}(q)=\frac{-ig^2}{q^2+(u\cdot
  q)^2-m^2}(2M\delta^{rr'})(2M\delta^{ss'}) ,
\eea
where the delta symbols correspond to the spins, and
$u^b=\frac{a^b}{\sqrt{M}}$.  The Breit effective potential is found via Fourier
transform of  the spacial part:
\bea
U(\vec{r})=-g^2\int\frac{d^d\vec{q}}{(2\pi)^{d}}\frac{
  e^{i\vec{q}\cdot \vec{r}}}{
  \vec{q}^2-(\vec{u}\cdot\vec{q})^2+m^2}.
\eea
Here $d=D-1$.
To perform the integration, we diagonalize the matrix in the denominator via an
appropriate rotation of the coordinates. Afterwards, this expression takes
the form  (cf. \cite{Alts})
\bea
U(\vec{r})=-g^2\int\frac{d^d\vec{q}}{(2\pi)^d}\frac{
e^{i\vec{q}\cdot\vec{r}}}{
\sum_iS_{ii}q^2_i+m^2},
\eea
where $S_{ij}=\delta_{ij}-u_iu_j$. Let us carry out changes of
variables similar to \cite{Alts}, that is,
$\tilde{q}_i=\sqrt{S_{ii}}q_i$, $\tilde{r}_i=\frac{r_i}{\sqrt{S_{ii}}}$. We find
\bea
U(\vec{r})=-\frac{g^2}{\det S}\int\frac{d^d\tilde{q}}{(2\pi)^d}\frac{
e^{i\tilde{q}\cdot\tilde{r}}}{\tilde{q}^2+m^2},
\eea
where $\tilde{q}^2=\sum_i\tilde{q}_i^2$. This integral yields
\bea
U(\vec{r})=\frac{|m|^{d/2-1}g^2}{(2\pi)^{d/2}\sqrt{\det
    S}}\frac{K_{d/2-1}(m\tilde{r})}{|\tilde{r}|^{d/2-1}},
\eea
where $K_n(x)$ is the modified Bessel function. 
Suggesting the vectors $u_i$ to be small we find $\det S=1-\vec{u}^2$,
$(S^{-1})_{ij}=\delta_{ij}+u_iu_j$, so, we get
$\bar{r}=\sqrt{r^2+(u\cdot r)^2}$, and
\bea
U(\vec{r})=\frac{|m|^{d/2-1}g^2}{(2\pi)^{d/2}\sqrt{1-u^2}}\frac{K_{d/2-1}(m
  \sqrt{r^2+(u\cdot r)^2})}{(\sqrt{r^2+(u\cdot r)^2})^{d/2-1}}.
\eea
The asymptotics of the function $K_n(x)$ is
\bea
K_n(x)|_{x\to\infty}\simeq\sqrt{\frac{\pi}{2x}}e^{-x}
\eea
Thus, we find that at large distances, the effective potential
displays 
exponential decay with the distance in any space-time dimension: 
\bea
\label{vr}
U(\vec{r})|_{r\to
  \infty}=\frac{\sqrt{m}g^2}{4\pi^2\sqrt{1-u^2}}\sqrt{\frac{\pi}{2}}
\frac{e^{-m\sqrt{r^2+(u\cdot r)^2}}}{[r^2+(u\cdot r)^2]^{3/4}}.
\eea
At small distances, the leading term in the Breit potential, for
example, in the
case $d=4$ (that is, $D=5$) is
\bea
U(\vec{r})|_{r\to 0}=\frac{mg^2}{4\pi^2\sqrt{1-u^2}}\frac{1}{r^2+(u\cdot r)^2}.
\eea
In the case $d=3$ ($D=4$), $d/2-1=1/2$, and $K_{1/2}(x)=
\sqrt{\frac{\pi}{2x}}e^{-x}$ exactly. Hence, the expression (\ref{vr})
in this space-time dimension is valid both at small and large
distances. Finally, in the case $d=2$ ($D=3$), the expression
(\ref{vr}) is valid at large distances whereas at small ones, the
effective potential grows logarithmically with the distance.   
In all cases, the Breit potential is anisotropic. 
However, in the limit $\vec{u}\to 0$, this potential
reproduces  Yukawa's (notice that for large enough $\vec{u}$,
the Lorentz-breaking term cannot be treated as a small  perturbation).

Next, we study the one-loop effective potential for the aether model
(\ref{acae}), to which an arbitrary scalar potential of the scalar
field  $V(\phi)$ is
added. Following a common procedure, we can split the field $\phi$ into
the sum of the background, $\Phi$, and the  quantum, $\chi$, fields . The
quadratic action for $\chi$  looks like
\bea
\label{acae1}
S_2[\chi]=-
\frac{1}{2}\int d^Dx\chi[\pa_b\pa^b+m^2+\frac{1}{M}(a\cdot\pa)^2+ 
V^{\prime\prime}(\Phi)]\chi.
\eea
The corresponding one-loop effective action of the $\Phi$ background field is
\bea
\Gamma^{(1)}=
\frac{i}{2}{\rm
  tr}\ln[\pa_b\pa^b+m^2+\frac{1}{M}(a\cdot\pa)^2+V^{\prime\prime}(\Phi)]=
-\int d^Dx V^{(1)}_{eff}(\Phi).
\eea
To find the effective potential we carry out the Fourier transform and
the Wick rotation and treate the background field as a constant. As a
result, we get
\bea
 V^{(1)}_{eff}(\Phi)=-\frac{1}{2}\int\frac{d^Dk_E}{(2\pi)^D}\ln[k^2+m^2-
\frac{1}{M}(a\cdot k)^2+V^{\prime\prime}(\Phi)]
\eea
Due to symmetry properties, we can replace
$k_ak_b\to\frac{k^2}{D}\delta_{ab}$. Thus, we get
\bea
 V^{(1)}_{eff}(\Phi)=-\frac{1}{2}\int\frac{d^Dk_E}{(2\pi)^D} 
\ln[k^2(1-\frac{a^2}{DM})+m^2+V^{\prime\prime}(\Phi)],
\eea
which is equal to
\bea
 V^{(1)}_{eff}(\Phi)
=-\Gamma(-\frac{D}{2})\left(\frac{m^2+V^{\prime\prime}(\Phi)}{4\pi(1-u^
    2/D)} \right)^{D/2}.
\eea
Thus, we have shown that the one-loop effective potential is finite
for $D=5$ and $D=3$ within the dimensional regularization framework
and it is free of Lorentz symmetry breaking. In four dimensions,
however, it is divergent  being equal to
\bea
 V^{(1)}_{eff}(\Phi)=-\frac{1}{16\pi^2\epsilon}(m^2+V^{\prime\prime}(\Phi))^2+
\frac{1}{16\pi^2}\frac{(m^2+V^{\prime\prime}(\Phi))^2}{(1-u^2/D)^2} 
\ln\left(\frac{m^2+V^{\prime\prime}(\Phi)}{\mu^2}\right),
\eea
where the constants, including the explicit $u^a$ dependence, are
absorbed into  a redefinition of $\mu$, hence the one-loop effective
potential is Lorentz symmetric.

The complete one-loop corrected effective potential is
\bea
V_{eff}(\Phi)=V(\Phi)+ V^{(1)}_{eff}(\Phi).
\eea

Now let us obtain the aether term for the spinor field. To do it, we start
with the action (1) and obtain the two-point function of the spinor
field.

The corresponding Feynman diagram looks like

\vspace*{3mm}

\hspace{6.0cm}
\Lengthunit=1.2cm
\GRAPH(hsize=2){\dashlin(.5,0)\mov(1,0){\halfdashcirc(1)[u]\halfcirc(1)[d]}
\mov(1.5,0){\dashlin(.5,0)}
 \ind(13,0){\bullet}\ind(3,0){\bullet}
}

\vspace*{3mm}

Its contribution has the form
\bea
S^{sp}_2=g^2\bar{\psi}(-p)\int\frac{d^Dk}{(2\pi)^D}
\frac{\as(\ks+m)\as}{(k^2-m^2)[(k+p)^2-m^2]}\psi(p).
\eea
Simplifying the product of matrices, we find
\bea
S^{sp}_2=g^2\bar{\psi}(-p)\int\frac{d^Dk}{(2\pi)^D}
\frac{2(a\cdot k)\as-a^2\ks+m a^2}{(k^2-m^2)[(k+p)^2-m^2]}\psi(p).
\eea

Here, the mass of scalar and spinor fields are chosen to be equal for
simplicity, however, the case of different masses is treated in the
same way.

Then we employ Feynman
representation, carry out change of variables and Wick rotation 
and disregard $p^2$ in
the denominator. As a result, the leading term (which will produce the
contribution with no more than one derivative) looks like
\bea
S^{sp}_2=ig^2\bar{\psi}(-p)\int\frac{d^Dk_E}{(2\pi)^D}\int_0^1 dx
\frac{-2x(a\cdot p)\as+xa^2\ps+m a^2}{(k^2_E+m^2)^2}\psi(p).
\eea
Integrating over $d^4k_E$ and $dx$, one finds
\bea
S^{sp}_2=ig^2\Gamma(2-D/2)\bar{\psi}(-p)
\frac{-(a\cdot p)\as+\frac{1}{2}a^2\ps+m
  a^2}{(4\pi)^{D/2}(m^2)^{2-D/2}}
\psi(p).
\eea
One finds that the term proportional to $(a\cdot p)$ exactly
reproduces the aether term introduced in \cite{Carroll}. Returning to
the Minkowski space and to the coordinate representation, one finds
\bea
\label{spae}
S^{sp}_2=
g^2\frac{\Gamma(2-D/2)}{(4\pi)^{D/2}(m^2)^{2-D/2}}
\bar{\psi}\left\{-i\as(a\cdot \pa)+\frac{i}{2}a^2\ds+m a^2\right\}\psi.
\eea
Thus, we succeeded to generate the aether-like term for the spinor
field, that is, the first term in the expression above. It is clear
that this term is finite in an odd-dimensional space-time. The
relevant, 
Lorentz-breaking part of this term is
\bea
\label{spae1}
S^{sp}_2=
-i\alpha \bar{\psi}\as(a\cdot \pa)\psi,
\eea
where $\alpha=g^2\frac{\Gamma(2-D/2)}{(4\pi)^{D/2}(m^2)^{2-D/2}}$.
Following \cite{KostColl2}, one can write down the following generic
form of the aether-like term 
\bea
\label{spinact}
S^{sp}_2=\frac{i}{2}\bar{\psi}_A\left\{(c_Q)_{\mu\nu AB}+(c_U)_{\mu\nu
    AB}+(c_D)_{\mu\nu AB}\right\}
\gamma^{\mu}\stackrel{\leftrightarrow}{\partial^{\nu}}\psi_B.
\eea
To compare with our model, we should restrict ourselves to the case of
only one spinor field,that is, $A=B=1$. In this situation, 
disregarding the Lorentz-invariant terms proportional to $a^2$, 
we should identify 
$((c_Q)_{\mu\nu AB}+(c_U)_{\mu\nu AB}+(c_D)_{\mu\nu AB})=-\alpha a_{\mu}a_{\nu}$.

If we consider the spinor action with the additive aether term, that
is,
\bea
S^{sp}=\int d^Dx \bar{\psi}(i\ds-m-i\as(a\cdot\pa))\psi,
\eea
one can find that the dispersion relations corresponding to this
action look like
\bea
p^2+(a\cdot p)^2a^2+2(a\cdot p)^2=m^2.
\eea
One can see that for the time-like $a^b=(a_0,\vec{0})$, one finds
$E^2(1+a^2_0)^2=\vec{p}^2+m^2$, which is evidently consistent at any
$a_0$. At the space-like $a^b=(0,\vec{a})$, one finds
$E^2=\vec{p}^2+m^2+(a\vec{p})^2\vec{a}^2-2(\vec{a}\cdot \vec{p})^2$
whose behaviour can be nonphysical in a certain interval of
$|\vec{a}|$.

One should note that an attempt to generate the aether term via the  
alternative Lorentz-breaking coupling of scalar field to the spinor
one, that is, via the action
\bea
S'=\int d^Dx\bar{\psi}(i\ds-m-\bs-g\phi)\psi,
\eea
in four dimensions can be shown to give zero result in the one-loop 
approximation.

\section{Aether term in the electrodynamics}

We start with the model of the spinor field coupled
to the electromagnetic field in a Lorentz-breaking manner in three,
four and five space-time dimensions.
We start with the introduction of a generic form of the magnetic coupling of
the electromagnetic field to the spinor one, characterized by the
action
\bea
\label{edlb1}
S=\int d^Dx  \Big[\bar{\psi}(i\ds-m-
\tilde{\epsilon}^{abc}b_{a}F_{bc})\psi-\frac{1}{4}F_{ab}F^{ab}\Big],
\eea
where $\tilde{\epsilon}^{abc}$ is a matrix-valued object
antisymmetric with respect to its Lorentz indices and defined as 
$\tilde{\epsilon}^{abc}=\epsilon^{abc}$, in $D=3$; $\tilde{\epsilon}^{abc}\equiv
\epsilon^{abcd}
\gamma_{d}$, in $D=4$, and
$\tilde{\epsilon}^{abc}=\epsilon^{abcde}\sigma_{de}$, in $D=5$.
Here $b_{a}$ is a vector implementing the Lorentz symmetry breaking. 
We note that in the three- and five-dimensional space-times the
one-loop 
contribution will be finite within the framework of the dimensional 
regularization.

In this theory, after imposing the Feynman gauge via adding the usual
gauge fixing term
\bea
S_{gf}=\frac{1}{2}A_a\partial^a\partial^bA_b.
\eea
the following Feynman rules take place:

\vspace*{2mm}

\hspace{2cm}
\Lengthunit=1.2cm
\GRAPH(hsize=3){\Linewidth{.6pt}\dashlin(1,0)
\ind(15,0){\;\;\;\;\;\;\;\;\;\;\;\;=\frac{i(\ps+m)}{p^{2}-m^{2}}}
}
\hspace{.5cm}
\Lengthunit=1.2cm
\GRAPH(hsize=3){\ind(5,0){\bullet}\dashlin(1,0)\mov(.5,0){\wavelin(0,.7)}
\ind(18,0){\;\;\;\;\;\;\;\;\;=-g \tilde{\epsilon}^{abc}b_a,}
}
\hspace{1cm}
\Lengthunit=1.2cm
\GRAPH(hsize=3){\Linewidth{.6pt}\wavelin(1,0)\ind(11,0){\;\;\;\;\;\;\;\;\;\;\;
    \;=\frac{i}{p^{2}}}
}

\vspace*{2mm}

\noindent Thus, we  find the lowest order contribution to the
two-point vertex 
function of the electromagnetic field is given by the following diagram:

\vspace*{3mm}

\hspace{6.0cm}
\Lengthunit=1.2cm
\GRAPH(hsize=2){\wavelin(.5,0)\mov(1,0){\dashcirc(1)}
\mov(1.5,0){\wavelin(.5,0)}\ind(15,0){\bullet}\ind(5,0){\bullet}
}

\vspace*{3mm}

\noindent Here the wavy line is for the external $F_{mn}$ field, and
the dashed line is for the propagator of the spinor $\psi$ field.
In five dimensions, this diagram produces the following contribution
to the effective action:
\bea\label{t1e}
S_2(p)=\frac{g^2}{2}\epsilon^{abcde}\epsilon^{a'b'c'd'e'}b_aF_{bc}(p)b_{a'}
F_{b'c'}(-p)\int\frac{d^5k}{(2\pi)^5}{\rm
  tr}\bigl[\sigma_{de}\,S(k)
\sigma_{d'e'}\,S(k+p)]
\eea
Thus, we can write down the following explicit expression for the new, 
Lorentz-breaking contribution to the quadratic effective action of the theory: 
\bea\label{i1a}
S_2(p)&=&-\frac{g^2}{2}\epsilon^{abcde}\epsilon^{a'b'c'd'e'}b_aF_{bc}(p)b_{a'}
F_{b'c'}(-p)\int\frac{d^5k}{(2\pi)^5}{\rm
  tr}\big[\sigma_{de}(\ks+m)
\sigma_{d'e'}(\ks+\ps+m)\big]\times\nonumber\\&\times&
\frac{1}{[k^2-m^2][(k+p)^2-m^2]}.
\eea
Here the matrices are defined just as in the previous section. 

The aether-like terms do not involve derivatives of $F_{ab}$, hence we
can impose condition $p_a=0$ in the internal lines from the very
beginning. As a result, we have
\bea
S_2(p)&=&-\frac{g^2}{2}\epsilon^{abcde}\epsilon^{a'b'c'd'e'}b_aF_{bc}(p)b_{a'}
F_{b'c'}(-p)
\int\frac{d^5k}{(2\pi)^5}\frac{1}{[k^2-m^2]^2}{\rm
  tr}\big[m^2\sigma_{de}\sigma_{d'e'}+\nonumber\\&+&k^mk^n\gamma_m\sigma_{de}
\gamma_n
\sigma_{d'e'}\big].
\eea
Here, we disregarded the terms proportional to a trace of an odd
number of gamma matrices, (which is either zero or produces irrelevant terms).
This expression can be rewritten as
\bea
S_2(p)=S_2^{(1)}(p)+S_2^{(2)}(p),
\eea
where
\bea
S_2^{(1)}(p)&=&-\frac{g^2m^2}{2}\epsilon^{abcde}\epsilon^{a'b'c'd'e'}
b_aF_{bc}(p)b_{a'}F_{b'c'}(-p)
{\rm tr}\big[\sigma_{de}\sigma_{d'e'}\big]
\times\nonumber\\&\times&
\int\frac{d^5k}{(2\pi)^5}\frac{1}{[k^2-m^2]^2},
\eea
\bea
S_2^{(2)}(p)&=&-\frac{g^2}{2}\epsilon^{abcde}\epsilon^{a'b'c'd'e'}b_aF_{bc}(p)
b_{a'}F_{b'c'}(-p)
{\rm tr}\big[\gamma_m\sigma_{de}\gamma_n\sigma_{d'e'}\big]
\times\nonumber\\&\times&
\int\frac{d^5k}{(2\pi)^5}\frac{k^mk^n}{[k^2-m^2]^2}.
\eea

By applying the relation (\ref{trg}) and using the definition 
$\sigma_{ab}=\frac{i}{2}[\gamma_a,\gamma_b]$, we get
\bea
{\rm tr}(\sigma_{de}\sigma_{d'e'})=
4(\eta_{dd'}\eta_{ee'}-\eta_{de'}\eta_{d'e}).
\eea

As a result, taking into account the relation
\bea
&&\epsilon^{abcde}\epsilon^{a'b'c'd'e'}\eta_{dd'}\eta_{ee'}\nonumber\\&=&
2(\eta^{aa'}\eta^{bb'}\eta^{cc'}-\eta^{aa'}\eta^{bc'}\eta^{cb'}+\eta^{ab'}
\eta^{bc'}\eta^{ca'}-\eta^{ac'}\eta^{bb'}\eta^{ca'}+
\eta^{ac'}\eta^{ba'}\eta^{cb'}-\eta^{ab'}\eta^{ba'}\eta^{cc'}),
\eea
and, consequently,
\bea
&&\epsilon^{abcde}\epsilon^{a'b'c'd'e'}\eta_{dd'}\eta_{ee'}b_aF_{bc}b_{a'}
F_{b'c'}=-8b^aF_{ac}b_bF^{bc}\equiv -8(b^aF_{ab})^2,
\eea
we can calculate the trace in the first term, so
\bea
\label{s21}
S_2^{(1)}(p)|_{p=0}=-2\frac{g^2i}{\pi^2}|m|^3(b^aF_{ab})^2.
\eea

Also, we can find the integral in $S_2^{(2)}$:
\bea
\int\frac{d^5k}{(2\pi)^5}\frac{k^ak^b}{[k^2-m^2]^2}=
-i\eta^{ab}\frac{|m|^3}{48\pi^2},
\eea
thus,
\bea
S_2^{(2)}(p)|_{p=0}&=&-\frac{g^2i|m|^3}{96\pi^2}\epsilon^{abcde}
\epsilon^{a'b'c'd'e'}b_aF_{bc}b_{a'}F_{b'c'}
{\rm tr}\big[\gamma_m\sigma_{de}\gamma^m\sigma_{d'e'}\big].
\eea
It remains to find ${\rm
  tr}\big[\gamma_m\sigma_{de}\gamma^m\sigma_{d'e'}\big]$. One can find
that ${\rm tr}\big[\gamma_m\sigma_{de}\gamma^m\sigma_{d'e'}\big]=
{\rm tr}\big[\sigma_{de}\sigma_{d'e'}\big]$. 
Substituting this result to the above expression, we find
\bea
\label{s22}
S_2^{(2)}(p)|_{p=0}=\frac{2g^2i}{3\pi^2}|m|^3(b^aF_{ab})^2.
\eea
 The final five-dimensional result being the sum of (\ref{s21}) and 
(\ref{s22}), after returning to the Minkowski space, is
\bea
S_2(p)|_{p=0}=-\frac{4g^2}{3\pi^2}|m|^3(b^aF_{ab})^2.
\eea

Let us consider other dimensions. In three space-time dimensions the
only way to apply this approach (without use of the derivative
expansion) is based on  the action
\bea
\label{edlb3}
S=\int d^3x\bar{\psi}(i\ds-m-g\epsilon^{abc}b_aF_{bc})\psi.
\eea
The corresponding correction is
\bea
S_2(p)&=&-\frac{g^2}{2}\epsilon^{abc}\epsilon^{a'b'c'}
b_aF_{bc}(p)b_{a'}F_{b'c'}(-p)\int\frac{d^3k}{(2\pi)^3}
\frac{1}{[k^2-m^2]^2}{\rm
  tr}\big[m^2+\nonumber\\&+&k^mk^n\gamma_m\gamma_n\big].
\eea
Calculating the trace and carrying out the Wick rotation, one finds
\bea
S_2(p)&=&-ig^2\epsilon^{abc}\epsilon^{a'b'c'}
b_aF_{bc}(p)b_{a'}F_{b'c'}(-p)\int\frac{d^3k}{(2\pi)^3}
\frac{-k^2+m^2}{[k^2+m^2]^2}.
\eea
After a straightforward calculation of the integral we find
\bea
S_2(p)&=&-ig^2\frac{|m|}{2\pi}\epsilon^{abc}\epsilon^{a'b'c'}
b_aF_{bc}(p)b_{a'}F_{b'c'}(-p).
\eea
We can multiply the Levi-Civita symbols which, after returning to the
Euclidean space, yields
\bea
S_2(p)&=&4g^2\frac{|m|}{2\pi}(b^aF_{ab})^2.
\eea

Finally, let us consider the four-dimensional space. In this case, the
spinor matter is coupled to the electromagnetic field via the action
\bea
\label{edlc}
S=\int d^4x\bar{\psi}(i\ds-m-g\epsilon^{abcd}b_aF_{bc}\gamma_d)\psi,
\eea
and the
Lorentz-breaking contribution to the quadratic effective action of the
electromagnetic field looks like
\bea\label{t1e1}
S_2(p)=\frac{g^2}{2}\epsilon^{abcd}\epsilon^{a'b'c'd'}b_aF_{bc}(p)b_{a'}
F_{b'c'}(-p)\int\frac{d^4k}{(2\pi)^4}{\rm
  tr}\bigl[\gamma_d\,S(k)\gamma_{d'}\,
S(k+p)]
\eea
The explicit form of this expression, at $p=0$, is
\bea
S_2(p)&=&-\frac{g^2}{2}\epsilon^{abcd}\epsilon^{a'b'c'd'}b_aF_{bc}(p)b_{a'}
F_{b'c'}(-p)
\int\frac{d^4k}{(2\pi)^4}\frac{1}{[k^2-m^2]^2}{\rm tr}
\big[m^2\gamma_d\gamma_{d'}+\nonumber\\&+&k^mk^n\gamma_m\gamma_d\gamma_n
\gamma_{d'}\big].
\eea
Proceeding with calculation of the trace, we arrive at
\bea
S_2(p)&=&-2g^2\epsilon^{abcd}\epsilon^{a'b'c'd'}b_aF_{bc}(p)b_{a'}
F_{b'c'}(-p)
\int\frac{d^4k}{(2\pi)^4}\frac{1}{[k^2-m^2]^2}
\big[m^2\eta_{dd'}+\nonumber\\&+&k^mk^n(\eta_{md}\eta_{nd'}-\eta_{mn}\eta_{dd'}
+\eta_{md'}\eta_{nd})\big].
\eea
We employ the relation $k^mk^n\to\frac{1}{4}\eta^{mn}k^2$.
After Wick rotation we find
\bea
S_2(p)&=&-ig^2\epsilon^{abcd}\epsilon^{a'b'c'd'}
b_aF_{bc}(p)b_{a'}F_{b'c'}(-p)\eta_{dd'}\int\frac{d^4k}{(2\pi)^4}
\frac{k^2+2m^2}{[k^2+m^2]^2}.
\eea
Despite the integral over $k$ is quadratically divergent, within the
dimensional regularization it yields finite result:
\bea
\int\frac{d^{4-\epsilon}k}{(2\pi)^{4-\epsilon}}
\frac{k^2+2m^2}{[k^2+m^2]^2}=\frac{m^{2(1-\epsilon/2)}}{(4\pi)^{2-\epsilon/2}}
[\Gamma(\epsilon/2)+\Gamma(-1+\epsilon/2)]=-\frac{m^2}{16\pi^2}+O(\epsilon).
\eea
Collecting all together, using the properties of product of the
Levi-Civita symbols and returning to the Minkowski space, we arrive at
\bea
\label{s2d4}
S_2(p)&=g^2\frac{m^2}{4\pi^2}(b^aF_{ab})^2.
\eea
We note that, had we done choice $k^mk^n\to\frac{1}{d}\eta^{mn}k^2$
with the subsequent choice $d=4-\epsilon$ as it has been done in
\cite{YM}, we 
would obtain the same
result (\ref{s2d4}). Therefore, differently from the Lorentz-breaking
Chern-Simons term \cite{list,YM}, this result appears to be
unambiguously 
determined.

As a result, within the dimensional regularization scheme 
the aether-like term is finite in all dimensions from three to five.
Its generic form is
\bea
L_{aether}=c(b^aF_{ab})^2,
\eea
where $c$ is a some constant depending on the space-time
dimension. This term is equivalent to the term introduced in 
\cite{KostColl2}:
\bea
L_{even}=-\frac{1}{4}(k_{F})_{\kappa\lambda\mu\rho}
F^{\kappa\lambda}F^{\mu\rho},
\eea
with the constant parameters $(k_F)_{\kappa\lambda\mu\rho}, c, b_a$ 
related in the following way:
$(k_{F})_{\kappa\lambda\mu\rho}=-c(\eta_{\kappa\mu}b_{\lambda}b_{\rho}+
\eta_{\lambda\rho}b_{\kappa}b_{\mu}-
\eta_{\kappa\rho}b_{\mu}b_{\lambda}+\eta_{\lambda\mu}b_{\kappa}b_{\rho})$.

Thus,  we have  succeeded to generate an effective Lagrangian with the
aether-like term, which has the form
\bea
\label{1}
L=-\frac{1}{4}F_{ab}F^{ab}+\frac{1}{2}u^au^b\eta^{cd}F_{ac}F_{bd},
\eea
{\bf where $u^a=\sqrt{|c|}b^a$.}
The equations of motion for this model are
\bea
\pa_aF^{ab}+2u^au^c\pa_aF_{\phantom{b}c}^b=0.
\eea
The energy-momentum tensor for this theory is evidently conserved due
to the 
homogeneity of the space-time, being equal to 
\bea
T^a_b=\frac{\pa L}{\pa(\pa_aA_c)}\pa_bA_c-L\delta^a_b,
\eea
which in this case is
\bea
T^a_b=(F^{ac}-2u^au^dF_d^c)\pa_bA_c-\delta^a_b(-\frac{1}{4}F_{cd}F^{cd}+
\frac{1}{2}u^mu^n\eta^{cd}F_{mc}F_{nd}).
\eea
However, due to the preferencial direction in the space-time
introduced by the 
$u^a$ vector, the angular momentum is not conserved.

Let us obtain the propagator in the model (\ref{1}).
In the Feynman gauge, the action takes the form
\bea
S=\int d^4x\frac{1}{2}A_a\Big(\eta^{ab}\pa^2-u^au^b\pa^2-\eta^{ab}
(u\cdot\partial)^2+(u^a\pa^b+u^b\pa^a)(u\cdot \pa)
\Big)A_b.
\eea
 We can  obtain the inverse operator for
\bea
\Delta^{ab}=\eta^{ab}\pa^2-u^au^b\pa^2-\eta^{ab}(u\cdot\partial)^2+(u^a\pa^b
+u^b\pa^a)(u\cdot \pa).
\eea
The most convenient way to do that is the indetermined coefficients
method. We 
suggest
\bea
\Delta^{-1}_{bc}=k_1\eta_{bc}+k_2\pa_b\pa_c+k_3u_bu_c+k_4u_b\pa_c+k_5u_c\pa_b,
\eea
with $\Delta^{ab}\Delta^{-1}_{bc}=\delta^b_c$.
It is clear that $k_1=[\pa^2-(u\cdot\partial)^2]^{-1}$. Then,
\bea
k_2&=&\frac{k_1(u\cdot\pa)^2}{\pa^4(1-u)^2}=[\pa^2-(u\cdot\partial)^2]^{-1}
\frac{(u\cdot\pa)^2}{\pa^4(1-u)^2},\nonumber\\
k_3&=&\frac{k_1}{1-u^2}=[\pa^2-(u\cdot\partial)^2]^{-1}\frac{1}{1-u^2},
\nonumber\\
k_4&=&-\frac{k_1(u\cdot\partial)}{\pa^2(1-u^2)}=[\pa^2-
(u\cdot\partial)^2]^{-1}\frac{(u\cdot\partial)}{\pa^2(1-u^2)},\nonumber\\
k_5&=&-\frac{k_1(u\cdot\partial)}{\pa^2(1-u^2)}=-[\pa^2-
(u\cdot\partial)^2]^{-1}\frac{(u\cdot\partial)}{\pa^2(1-u^2)}.
\eea
The dispersion relations can be found from the poles of the
denominators ${[\pa^2-(u\cdot\partial)^2]}^{-1}$ and
$({\pa^2})^{-1}$. The second one is the usual Lorentz-invariant one,
whereas the first one yields the following cases for $(+-\ldots-)$ signature:

\noindent (i) The $u^a$ is space-like. We get
$E^2=\vec{p}^2-(\vec{u}\cdot\vec{p})^2$. This possibility
is consistent only for small $|\vec{u}|$.\\
(ii) The $u^a$ is time-like. We get $E^2(1+u^2_0)=\vec{p}^2$. This
case is consistent for any $u_0$.


It is interesting also to generate the aether-like term for the spinor
field from its coupling with the electromagnetic field (\ref{edlb1}). 
The lowest-order Feynman diagram looks like

\vspace*{3mm}

\hspace{6.0cm}
\Lengthunit=1.2cm
\GRAPH(hsize=2){\dashlin(.5,0)\mov(1,0){\halfdashcirc(1)[u]\halfwavecirc(1)[d]}
\mov(1.5,0){\dashlin(.5,0)}
 \ind(13,0){\bullet}\ind(3,0){\bullet}
}

\vspace*{3mm}

Its contribution is
\bea
I=4g^2\int\frac{d^Dk}{(2\pi)^D}\bar{\psi}(-p)\tilde{\epsilon}^{abc}
[\gamma^{d}(k_{d}+p_{d})-m]\tilde{\epsilon}^{a'b'c'}
\eta_{cc'}b_{a}b_{a'}
\frac{k_{b}k_{b'}}{k^2[(k+p)^2-m^2]}\psi(p).
\eea
Then we use the Feynman representation which allows us to write
\bea
I&=&4g^2\int\frac{d^Dk}{(2\pi)^D}\int_0^1dx\bar{\psi}(-p)\tilde{\epsilon}^{abc}
\eta_{cc'}b_{a}b_{a'}
[\gamma^{d}(k_{d}+p_{d}(1-x))-m]\tilde{\epsilon}^{a'b'c'}\times\nonumber\\
&\times&
\frac{(k_b-p_bx)(k_{b'}-p_{b'}x)}{[k^2+p^2x(1-x)-m^2x]^2}\psi(p).
\eea
Afterwards, let us again restrict ourselves by the terms involving at
most first order in external momenta $p_a$ (that is, the terms whose
form is similar
to those ones from the Dirac Lagrangian). Moreover, the term 
proportional to $m$ can be easily shown to generate only
Lorentz-invariant contribution proportional to $b^ab_a$
(indeed, this term does not contain explicit momenta $p^a$,
hence the $b^a$ can be contracted only among themselves), hence
we disregard this term and also neglect $p^2$ in the denominator and rest with 
\bea
I&=&-4g^2\int\frac{d^Dk}{(2\pi)^D}\int_0^1dx\bar{\psi}(-p)
\tilde{\epsilon}^{abc}
\eta_{cc'}b_{a}b_{a'}\times\nonumber\\&\times&
\frac{(\gamma^dk_dk_bp_{b'}x+\gamma^dk_dk_{b'}p_bx-\gamma^dp_d(1-x)k_bk_{b'})
\tilde{\epsilon}^{a'b'c'}}{[k^2-m^2x]^2}\psi(p).
\eea
Then, we replace $k_ak_b\to\eta_{ab}\frac{k^2}{D}$ and carry out the
Wick 
rotation $k_0\to ik_{0E}$ (hence, $k^2\to k^2_E$), as a result we arrive at
\bea
I&=&\frac{4i}{D}g^2\int\frac{d^Dk_E}{(2\pi)^D}\frac{k^2_E}{[k^2_E+m^2x]^2}
\int_0^1dx\bar{\psi}(-p)\tilde{\epsilon}^{abc}
\eta_{cc'}b_{a}b_{a'}\times\nonumber\\&\times&
(\gamma_{b}p_{b'}x+\gamma_{b'}p_bx-\gamma^dp_d(1-x)\eta_{bb'})
\tilde{\epsilon}^{a'b'c'}\psi(p).
\eea
Integrating over $k$ together with the subsequent returning to the
Minkowski 
space, we find
\bea
\label{ispex0}
I&=&\frac{2g^2}{(4\pi)^{D/2}}|m|^{D-2}\Gamma(1-\frac{D}{2})\bar{\psi}(-p)
\tilde{\epsilon}^{abc}
\eta_{cc'}b_{a}b_{a'}\times\nonumber\\&\times&
\int_0^1dx
x^{D/2-1}(\gamma_{b}p_{b'}x+\gamma_{b'}p_bx-\gamma^dp_d\eta_{bb'}
(1-x))\tilde{\epsilon}^{a'b'c'}\psi(p).
\eea
Afterwards, we integrate over $x$ and arrive at
\bea
\label{ispex}
I&=&\frac{2g^2}{(4\pi)^{D/2}}|m|^{D-2}\Gamma(1-\frac{D}{2})\bar{\psi}(-p)
\tilde{\epsilon}^{abc}
\eta_{cc'}b_{a}b_{a'}\times\nonumber\\&\times&
\left[\frac{2}{D}(\gamma_{b}p_{b'}+\gamma_{b'}p_b)-(\frac{2}{D}-\frac{2}{2+D})
\gamma^dp_d\eta_{bb'}\right]\tilde{\epsilon}^{a'b'c'}\psi(p).
\eea

It remains to simplify the matrix products.
For example, at $D=3$ we have
\bea
I&=&\frac{2g^2}{3\pi}|m|
\bar{\psi}(-p)\gamma^{a}p^{b}b_{a}b_{b}\psi(p),
\eea
which in the coordinate representation yields
\bea
I=\frac{2i|m|g^2}{3\pi}\bar{\psi}\bs(b\cdot \pa)\psi,
\eea
that is, the aether-like term of the same form as in (\ref{spae1}). 
The calculations at $D=4,5$ do not principially differ. So, in $D=4$
the 
contribution (\ref{ispex}) diverges being, at $D=4+\epsilon$, of the form
\bea
I= \frac{7 im^2g^2}{24\pi^2\epsilon}\bar{\psi}\bs(b\cdot \pa)\bs\psi+{\rm fin},
\eea
which again reproduces the form (\ref{spae1}), however, for the 
renormalizability such a term should present in the Lagrangian from
the very beginning.

Finally, one can show that at $D=5$ the term (\ref{spae}) also arises,
and it 
is finite in the dimensional regularization framework. Calculating the 
products of matrices in (\ref{ispex}) in this case, we find
\bea
\label{ispex1}
I&=&-\frac{484g^2}{105\pi^2}i|m|^3\bar{\psi}\bs(b\cdot\pa)\psi.
\eea
Therefore, we succeeded in generating the aether-like term also in this
case, and it is finite within the dimensional regularization
prescription. 
By analogy with the case of the coupling with the scalar field, one
can write down the one-loop quantum contribution to the aether term
for the spinor field 
in all space-time dimensions as
\bea
I=i\tilde{\alpha}\bar{\psi}\bs(b\cdot \pa)\psi,
\eea
where $\tilde{\alpha}$ is a some constant depending on the dimension
of the 
space-time.  
 Comparing this term with the standard form of the aether-like action
 for the spinors (\ref{spinact}) originally introduced in
 \cite{KostColl2}, we conclude that the constant parameters in
 \cite{KostColl2} and those used in this paper are related as 
$((c_Q)_{\mu\nu AB}+(c_U)_{\mu\nu AB}+(c_D)_{\mu\nu AB})=-
\tilde{\alpha} b_{\mu}b_{\nu}$.

\section{Summary}

We succeeded in generating CPT-even Lorentz-breaking models with
aether-like terms for the scalar and vector fields through appropriate
couplings to the spinor matter and studied their different physical
aspects. We also generated aether-like terms for the spinor field from
the same interactions. 
We have shown that the aether-like terms,
arising as perturbative corrections, are finite within the dimensional
regularization framework in three and five
dimensions, and in the case of the electrodynamics -- also in four
dimensions, which is highly nontrivial since this contribution is
superficially quadratically divergent. We note that the last result is
unambiguously 
determined
unlike the four-dimensional Chern-Simons term \cite{JK,list,YM}
which reflects naturally the fact that the four-dimensional
Chern-Simons term is related with the ABJ anomaly \cite{YM} whereas
there is no known analog of this anomaly for the aether term. 

We have calculated the tree-level Breit effective potential of
scalar-spinor coupling which was shown to be anisotropic, as well as
the one-loop effective potential for the scalar field which is shown
to be free of Lorentz symmetry breaking. We also studied the
dispersion relations for the modified scalar, spinor and
electromagnetic 
field models.

The natural continuation of this study would consist in the
implementation of the condition of compactness for the extra dimension
and the study of the physical impacts of the extra compact dimensions
for these terms (some studies for the phenomenological aspects of the
compact extra dimensions can be found in \cite{aether}). We are
planning to carry out this study in a forthcoming 
paper.

{\bf Acknowledgements.}  This work was partially supported by Conselho
Nacional de Desenvolvimento Cient\'{\i}fico e Tecnol\'{o}gico (CNPq),
Funda\c{c}\~{a}o de Amparo \`{a} Pesquisa do Estado de S\~{a}o
Paulo (FAPESP), Coordena\c{c}\~{a}o de Aperfei\c{c}oamento do Pessoal
do Nivel Superior (CAPES: AUX-PE-PROCAD 579/2008) and
CNPq/PRONEX/FAPESQ. A. Yu. P. has been supported by the CNPq project 
303461-2009/8.

\end{document}